\documentclass[12pt]{article}
 \pdfoutput=1
 \usepackage{graphicx}
  \usepackage{epsfig}
  \usepackage{amsmath}
  \usepackage{amssymb}
  \usepackage{color}
  \newlength{\abstractwidth}
  \newcommand{\be}{\begin{equation}}
  \newcommand{\ee}{\end{equation}}
  
  \renewcommand{\title}[1]{\vbox{\center\bf{\Large{#1}}}\vspace{5mm}}
  \renewcommand{\author}[1]{\vbox{\center#1}\vspace{5mm}}
  \newcommand{\address}[1]{\vbox{\center\em#1}}
  \newcommand{\email}[1]{\vbox{\center\tt#1}\vspace{5mm}}

\begin{document}

\begin{titlepage}
\rightline{SU-ITP-13/08}
\begin{center}
\hfill \\
\hfill \\
\vskip 1cm

\title{Black holes and the butterfly effect}

\author{Stephen H. Shenker and Douglas Stanford}

\address{
 Stanford Institute for Theoretical Physics {\it and} \\
 Department of Physics, Stanford University \\
 Stanford, CA 94305 USA \\ \bigskip 
 Kavli Institute for Theoretical Physics \\
 University of California \\
 Santa Barbara, CA 93106-4030 USA
}

\email{sshenker@stanford.edu,
salguod@stanford.edu}

\end{center}
  
  \begin{abstract}
  We use holography to study sensitive dependence on initial conditions in strongly coupled field theories. Specifically, we mildly perturb a thermofield double state by adding a small number of quanta on one side. If these quanta are released a scrambling time in the past, they destroy the local two-sided correlations present in the unperturbed state. The corresponding bulk geometry is a two-sided AdS black hole, and the key effect is the blueshift of the early infalling quanta relative to the $t = 0$ slice, creating a shock wave. We comment on string- and Planck-scale corrections to this setup, and discuss points that may be relevant to the firewall controversy.
  \end{abstract}

  \end{titlepage}

\tableofcontents

\baselineskip=17.63pt

\section{Introduction}

Entanglement is a  central property of quantum systems.  It plays a crucial role in the theory of quantum information, quantum many body systems and quantum field theory.   Two subsystems A  and B of a quantum system are entangled in the state $| \psi \rangle$  if the total Hilbert space ${ \cal H}$ can be decomposed into subfactors, ${\cal H} = {\cal H}_A \otimes {\cal H}_B$  and the density matrix $\rho_A$ obtained by tracing out ${\cal H}_B$, $\rho_A = tr_{{\cal H}_B} [|\psi\rangle\langle \psi|]$,  is not pure.   This can be diagnosed using the von Neumann entropy $S_A = - tr_{{\cal H}_A }[\rho_A \log \rho_A]$ which is greater than zero if and only if $|\psi \rangle$ is entangled.    

Entropy of entanglement of the ground state can be used as a diagnostic of topological order in gapped quantum systems \cite{Kitaev:2005dm}.   In conformal quantum field theories (CFTs) defined on a sphere, the entropy of entanglement  between hemispheres of the vacuum state  has been shown to be  the correct measure of the number of degrees of freedom which decreases under renormalization group flow, encompassing the $c$, $a$ and $F$ theorems \cite{Myers:2010xs,Casini:2011kv}.

Entanglement in highly excited states is also of great importance.    If $|\psi\rangle$ is a typical state and A is a small subsystem then $\rho_A$ describes a thermal distribution.   B serves as a heat bath for A.     An exactly thermal density matrix can be obtained from a pure entangled state using the thermofield double construction.  Consider two identical subsystems, L and R.  Write a pure state $| \Psi \rangle$ in the total Hilbert space:

\be
|\Psi \rangle = \frac{1}{Z^{1/2}}\sum_n e^{-\beta E_n/2}~ |n\rangle_L |n \rangle_R\label{tfd}.
\ee
Tracing over the R Hilbert space leaves a precisely thermal density matrix for the L system:
\be
\rho_L = \frac{1}{Z} \sum_n e^{-\beta E_n} |n\rangle \langle n |.
\ee

These ideas have a holographic realization in the AdS/CFT correspondence \cite{Maldacena:2001kr}.    If the L and R systems are CFTs with AdS duals, and the temperature is sufficiently high, then $|\Psi\rangle$ describes a large eternal AdS Schwarzschild black hole with Hawking temperature $T_H = 1/\beta$.  In this context $|\Psi\rangle$ is referred to as the Hartle-Hawking state. The UV degrees of freedom of the L and R CFTs describe dynamics at the disconnected large radius asymptotic regions of the eternal black hole geometry. The entropy of entanglement $S_L = - tr [\rho_L \log \rho_L]$ is the Bekenstein-Hawking entropy of the black hole given by the area of the event horizon, $S_L = A_h/4G_N$.  

Entanglement entropy has a more general holographic interpretation.   It was proposed by Ryu and Takayanagi \cite{Ryu:2006bv} (RT) that the entanglement entropy of a region A in a CFT in a state $ |\psi\rangle$ is given by the area (in Planck units) of the minimal area codimension two spacelike surface  whose asymptotic  boundary is the boundary of A in the geometry dual to $|\psi\rangle$.    This proposal was first proved in the case of spherical boundaries in  \cite{Casini:2011kv} and recently explained in the most general static case in \cite{Lewkowycz:2013nqa}.  The RT proposal has been extended to nonstatic geometries in \cite{Hubeny:2007xt}.

Thermal systems share another basic property--chaos.   Starting from rather special states these systems  evolve to much more disordered typical states.   There is sensitive dependence on initial conditions, so that initially similar (but orthogonal) states evolve to be quite different.    In the subject of quantum information and black holes, such chaotic behavior has come to be referred to as ``scrambling," and it has been conjectured that black holes are the fastest scramblers in nature \cite{Hayden:2007cs,Sekino:2008he,Susskind:2011ap,Lashkari:2011yi}.  The time it takes such fast scramblers to render the density matrix of a small subsystem A essentially exactly thermal is conjectured to be $t \sim \beta \log S$ where  $S$ is the entropy of the system.

Scrambling can disrupt certain kinds of entanglement.  In particular, if the pattern of entanglement is characteristic of an atypical state,  scrambling, which takes the state toward typicality, can destroy it. This interplay is at the heart of the firewall proposal \cite{Almheiri:2012rt}.   These authors argue that the existence of a smooth region connecting the outside and inside of the horizon requires special entanglement of degrees of freedom on the two sides. But during the evaporation of the black hole the system scrambles, and these delicate correlations are destroyed.   No smooth region can remain.\footnote{A related argument was provided in \cite{Braunstein}, along with a claimed resolution that relies on a non-standard model of Hawking radiation.}

In this paper we will study the interplay of entanglement and scrambling using holographic tools, assuming the validity of the classical bulk geometry.  We will use a fine grained measure of the correlation between two subsystems called mutual information.   If A and B are subsystems then the mutual information $I$ is defined to be $I =S_A + S_B - S_{A \cup B}$.

This quantity has been studied holographically using RT surfaces in a number of papers. Mutual information and entanglement entropy have been used to diagnose thermalization after a quantum quench, in both conventional \cite{Calabrese:2005in,de2006entanglement} and holographic setups \cite{Hubeny:2007xt,AbajoArrastia:2010yt,Albash:2010mv,Balasubramanian:2010ce,Balasubramanian:2011at,Allais:2011ys,Das:2011nk,Nozaki:2013wia}. A common feature in the evolution of $I$ is a sharp transition in which the connected $A\cup B$ minimal surface exchanges dominance with the union of the disconnected $A$ and $B$ surfaces. At this point, $I$ goes to zero and stays there in a continuous but non differentiable way.\footnote{ When we say $I$ is zero we mean the coefficient of $\frac{1}{G_N}$, or in the large $N$ field theory context the coefficient of $N^2$, vanishes.  There will continue to be a nonzero value of subleading strength.}

Here we will focus on the eternal black hole setup discussed above, with regions A in the L system and B in the R system.   $I$ has been studied for this situation  in \cite{Morrison:2012iz,Hartman:2013qma}. The surface determining $S_{A \cup B}$ may pass behind the horizon, giving some information about that region.   In particular, Hartman and Maldacena \cite{Hartman:2013qma} studied $I$ between two regions as both boundary times are increased.\footnote{In this paragraph and the one below, we are referring to the physical time conjugate to $H_R + H_L$, which runs forwards on both CFTs. In the rest of the paper $t$ will refer to the Killing time, which is conjugate to $H_R - H_L$ and so runs forwards on the right CFT but backwards on the left.}   The Hartle-Hawking state $|\Psi\rangle$ is not invariant under this time evolution and $I$ rapidly decreases, going to zero linearly in a thermal time $\beta$.

Van Raamsdonk \cite{VanRaamsdonk} made the important point that while an arbitrary unitary transformation applied to the left handed CFT leaves the density matrix describing right handed CFT observables unchanged, it will change the relation between degrees of freedom on both sides and hence the geometry behind the horizon. Certain unitaries correspond to local operators, which can create a pulse of radiation propagating just behind the horizon \cite{Czech} which in some ways resembles a firewall \cite{VanRaamsdonk} .

The new feature that we will explore is sensitivity to a very small initial perturbation.   We imagine choosing  regions A and B in the L and R CFTs  at time $t =0$.  Because of the atypical local structure of entanglement in the thermofield double state, $A$ and $B$ may be highly entangled, even if they are small subsystems of $L$ and $R$. The state at an earlier time, $-t_w$ does not have these correlations but is carefully ``aimed" to give them at $t=0$.  We then consider the effect of injecting  a small amount of energy $E$ into the L system, by throwing a few quanta towards the horizon at time $-t_w$.  One expects that the CFTs dual to black holes have sensitive dependence on initial conditions, and this small perturbation should touch off chaotic behavior in the L theory, disturbing the careful aiming. The resulting Schrodinger picture state at $t=0$, $|\Psi'\rangle$, should be more typical than the thermofield double state. In particular, it should have less entanglement between A and B.

At first, this presents a puzzle: entanglement is determined by geometrical data, and, naively, the geometry is unaffected by the addition of a few quanta. However, the boundary time $t = 0$ defines a frame in the bulk, and relative to this frame, the quanta released a time $t_w$ in the past will have exponentially blue-shifted energy. Their backreaction must be included. The relevant bulk geometry can be described as a shock wave \cite{Dray:1985yt}, a limiting case of a Vaidya metric. Closely related configurations have been discussed in a context similar to ours by \cite{Silverstein}. \footnote{In particular  \cite{Silverstein} discussed, in the one sided black hole context, highly boosted horizon hugging branes.}  In the 3D BTZ case that we focus on, the dimensionless effect of the quanta on RT surfaces passing through the horizon at $t = 0$ is proportional to $\frac{E}{M} e^{2\pi t_w/\beta}$, where $M$ is the mass of the black hole. Eventually this effect becomes of order one, RT surfaces exchange dominance, and $I$ drops to zero.  This begins when $t_w$ becomes of order $t_* \sim \frac{\beta}{2\pi} \log \frac{M}{E}$. Assuming $E$ takes  the smallest reasonable value, the energy in one quantum at the Hawking temperature $E \sim T_H$, the time $t_*$ is
\be
t_* \sim \frac{\beta}{2\pi} \log S
\ee
which is the fast scrambling time. This is our central result. Flat space stringy effects will not change $t_*$. However, as we will emphasize in Section \ref{speffects}, we are unable to reliably exclude the possibility that stringy effects in the presence of the black hole will be parametrically stronger and lead to a smaller $t_*$.

The logarithmic behavior arises as in \cite{Sekino:2008he,Susskind:2011ap} from the relation between Rindler time evolution and Minkowski boosts. The connection between fast scrambling and large boosts has also been emphasized recently in \cite{MaldacenaSusskind}.   This importance of this time scale in black hole physics was pointed out in earlier work, including 
\cite{Schoutens:1993hu}.

The outline of our paper is as follows:
In Section \ref{qubit} we will illustrate the basic idea of scrambling destroying mutual information in a simple qubit system.  In Section \ref{holographic}  we will describe the basic geometrical constructions used and calculate the mutual information holographically, assuming Einstein gravity.  We also discuss correlation functions as probes of entanglement.  In Section \ref{speffects} we will address string- and Planck-scale corrections to the results from \S~\ref{holographic}. In Section \ref{discussion}, we will discuss various issues, including the connection to other notions of scrambling and the possible relevance to firewall ideas.

\section{A qubit model}\label{qubit}
Directly following the thermalization of a chaotic system is challenging, almost by definition. Our primary tool in this paper, holography, is powerful but somewhat indirect, and we would like to illustrate the effect of scrambling on entanglement in a simpler context. One tractable approach is to study a system with Haar random dynamics, which powerfully disrupt local two-sided mutual information. We pursue this in appendix \ref{random}. In the present section, we will consider a more physical system, by numerically evolving a collection of thermal qubits. Although we are limited to a rather small system, the basic effect will be visible.

Using sparse matrix techniques, it is possible to time-evolve pure states of twenty to thirty qubits. We will be less ambitious, studying a system ($L$) made up of ten qubits, plus another ten for the thermofield double ($R$). We will use an Ising Hamiltonian, with both transverse and parallel magnetic fields:
\be
H_L = \sum_{i=1}^{10} \Big\{\sigma_z^{(i)}\sigma_z^{(i+1)} - 1.05 \ \sigma_x^{(i)} + 0.5 \  \sigma_z^{(i)}\Big\}.
\ee
The coefficients -1.05 and 0.5 are chosen, following \cite{Hastings}, to ensure that the Hamiltonian is far from integrability.

\begin{figure}[ht]
\label{numerics}
\begin{center}
\includegraphics[scale = 0.45]{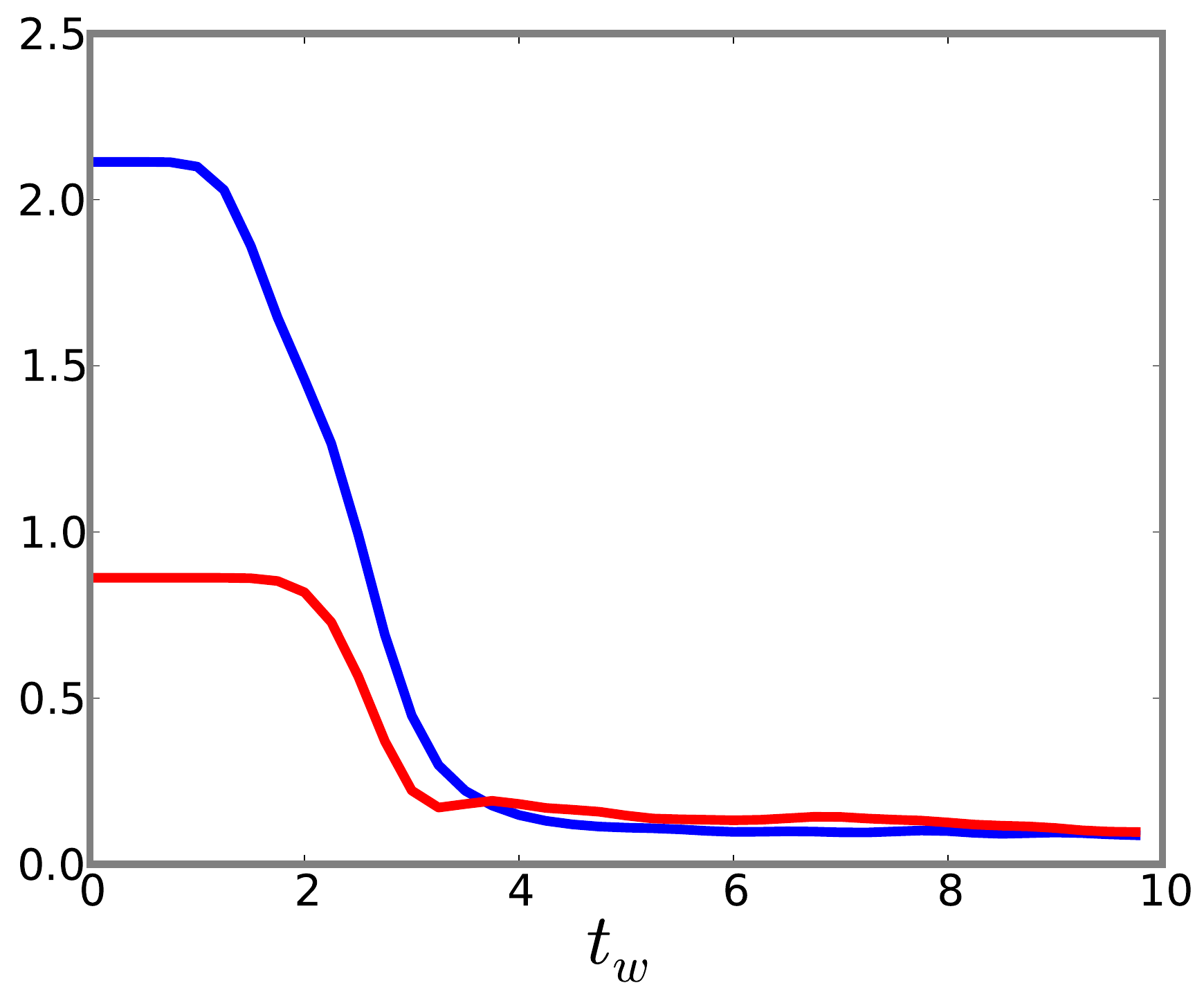}
\caption{Mutual information (upper, blue) and spin-spin correlation function (lower, red) in the perturbed state $|\Psi'\rangle$, as a function of the time of the perturbation $t_w$. The delay is a propagation effect; if the perturbation at site five is sufficiently recent, sites one and two are unaffected.}\label{numerical}
\end{center}
\end{figure}
Our procedure is to prepare the thermofield double state $|\Psi\rangle$, as in Eq.~\eqref{tfd}, at a reference time $t = 0$. We then apply a perturbation $\sigma_z^{(5,L)}$ to the fifth qubit of the $L$ system at a time $t_w$ in the past. In other words, we consider the perturbed state
\be
|\Psi'\rangle = e^{-iH_Lt_w}\sigma_z^{(5,L)}e^{iH_Lt_w}|\Psi\rangle.
\ee
Notice that the applied operator acts trivially on the $R$ system. In the state $|\Psi'\rangle$, we then compute the mutual information between sites one and two and their thermofield doubles. The result is the blue curve in Fig.~\ref{numerical}.

In the unperturbed state $|\Psi\rangle$, the mutual information is near-maximal. For small $t_w$, this continues to be true in the perturbed state. However, as $t_w$ increases and the perturbation is moved farther into the past, $I(A;B)$ drops sharply before leveling off at a floor value. By studying the same problem for eight or nine qubits instead of ten, we note that the floor of the mutual information appears to decrease with the total size of the system.

Although mutual information is a particularly thorough measure of $AB$ correlation, the same basic phenomenon is visible in simpler quantities. A useful example is the spin-spin two point function $\langle \Psi'| \sigma_z^{(1,L)}\sigma_z^{(1,R)}|\Psi'\rangle$, between spin one in the $L$ system and spin one in the $R$ system. This quantity is plotted as a function of $t_w$ in Fig.~\ref{numerical}, and we see that it exhibits the same qualitative behavior as the mutual information: the special local correlations of the thermofield double state are destroyed by a small perturbation applied sufficiently long in the past.

\section{A holographic model}\label{holographic}
In this section we will present our main result, a bulk geometry that illustrates the sensitivity of specific entanglements in the thermofield double state to mild perturbations long in the past. We will use RT surfaces and correlation function probes to analytically follow the loss of local correlation between the $L$ and $R$ sides. We will work with Einstein gravity in 2+1 bulk dimensions in this section, deferring comments about string- and Planck-scale effects to section \ref{speffects}, and deferring comments about higher dimensional Einstein gravity to appendix \ref{higherd}.

\subsection{Unperturbed BTZ}
Let us begin by reviewing the geometrical dual of the unperturbed thermofield double state of two CFTs \cite{Maldacena:2001kr}. This is an AdS-Schwarzschild black hole, analytically extended to include two asymptotically AdS regions. We think of the CFTs as living at the boundaries of the respective regions. In 2+1 bulk dimensions, the black hole solution is a BTZ metric, which can be presented as
\begin{align}
ds^2 = -\frac{r^2 - R^2}{\ell^2}dt^2 + \frac{\ell^2}{r^2-R^2}dr^2 + r^2 d\phi^2 \\
\phi\sim \phi+ 2\pi \hspace{20pt} R^2 = 8 G_N M\ell^2 \hspace{20pt} \beta = \frac{2\pi\ell^2}{R},\label{mass}
\end{align}
where we use $\ell$ to denote the AdS radius, and $R$ to denote the horizon radius. In what follows, it will often be more convenient to use Kruskal coordinates, which smoothly cover the maximally extended two-sided geometry.
\begin{figure}[ht]
\begin{center}
\includegraphics[scale = 0.6]{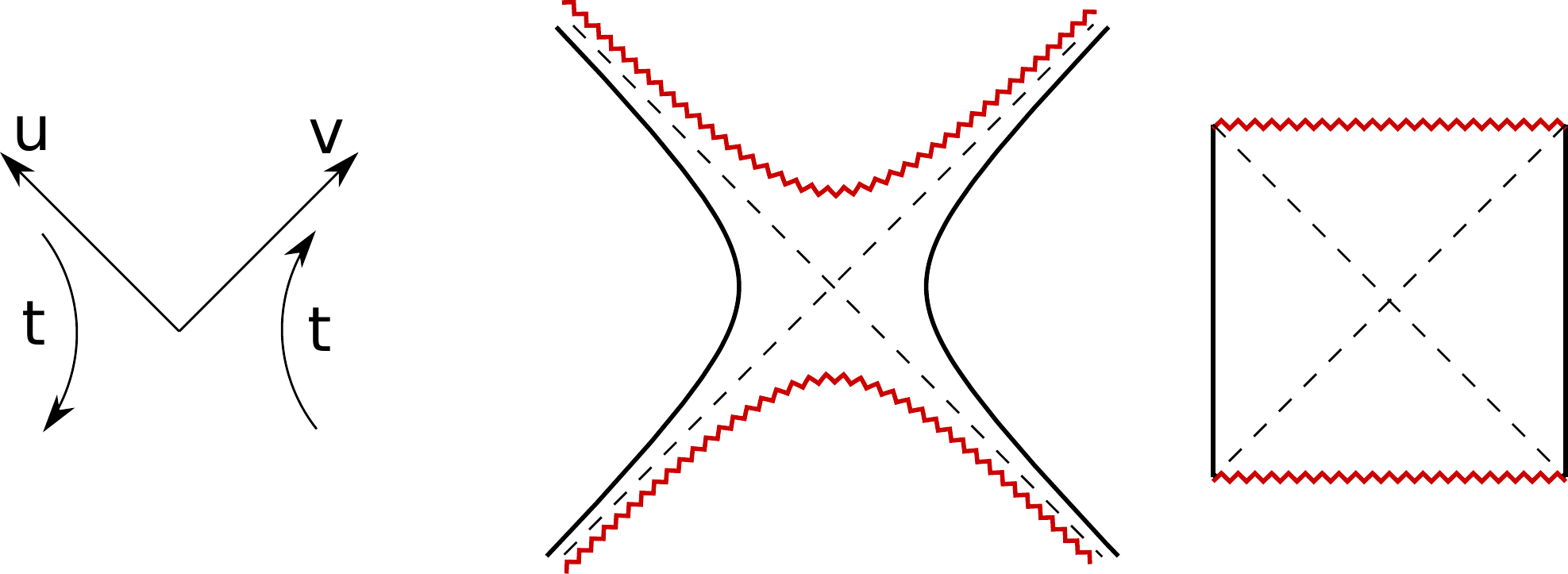}
\caption{The Kruskal diagram (center) and Penrose diagram (right) for the BTZ geometry.}\label{btzfig}
\end{center}
\end{figure}
In these coordinates, the metric is
\begin{align}
\label{kruskal}
ds^2 = \frac{-4 \ell^2 du dv + R^2(1-uv)^2d\phi^2}{(1+uv)^2}.
\end{align}
We will use the standard $u,v$ convention so that the right exterior has $u<0$ and $v>0$. The two boundaries are at $uv = -1$, and the two singularities are at $uv = 1$. 

Below, we will be interested in computing geodesic distances between points in the BTZ geometry. Since BTZ is a quotient of AdS, we can use the formula for geodesic distance in pure AdS${}_{2+1}$:
\be
\label{dist}
\cosh \frac{d}{\ell} = T_1 T_1' + T_2 T_2' - X_1 X_1'-X_2X_2',
\ee
where we've used the embedding coordinates
\begin{align}
T_1 &= \frac{v+u}{1+uv} = \frac{1}{R}\sqrt{r^2-R^2} \ \sinh \frac{R t}{\ell^2} \notag\\
T_2 &= \frac{1-uv}{1+uv}\cosh \frac{R\phi}{\ell} = \frac{r}{R} \ \cosh \frac{R\phi}{\ell}  \label{embed}\\
X_1 &= \frac{v-u}{1+uv} = \frac{1}{R}\sqrt{r^2-R^2} \ \cosh \frac{R t}{\ell^2} \notag \\
X_2 &= \frac{1-uv}{1+uv}\sinh \frac{R\phi}{\ell} = \frac{r}{R} \ \sinh \frac{R\phi}{\ell}. \notag
\end{align}
These coordinates also allow us to relate $(r,t)$ to $(u,v)$. Note, in particular, that the left asymptotic region can be reached in the $(r,t)$ coordinates by adding $i\beta/2$ to $t$.

\subsection{BTZ shock waves}\label{shock}
Having set up the bulk dual of the thermofield double state of the two CFTs, we would like to very mildly perturb it. As an example, we might add a few particles at the left boundary, and let them fall into the black hole. Naively, this would seem to have an insignificant effect on the geometry. However, as is familiar from Rindler space, translation in the Killing time $t$ is a boost in the $(u,v)$ coordinates, and if we release a perturbation with field theory energy $E$ from the boundary at a time $t_w$ long in the past,\footnote{We emphasize that $t$ is the Killing time coordinate. In our convention, it runs forward on the right boundary and backwards on the left (see Fig.~\ref{btzfig}). In particular, a perturbation released at time $t_w$ from the left boundary is in the past of the $t = 0$ slice if $t_w>0$.} it will cross the $t = 0$ slice with proper energy
\be E_p \sim \frac{E\ell}{R} e^{Rt_w/\ell^2}
\ee
as measured in the local frame of that slice. In this frame, the perturbation will be a high energy shock following an almost null trajectory close to the past horizon.

If $t_w$ is sufficiently large, we must include the backreaction of this energy. For the simplest case of spherically symmetric null matter, the backreacted metric is a special case of the AdS-Vaidya solution.\footnote{This metric corresponds to a boundary source adjusted to make all particles fall through the horizon at the same time.  We comment further on this choice in the Discussion.  The use of a classical  metric is simplest to justify if we consider a perturbation that corresponds to a large but fixed number of quanta in the small $G_N$ limit. However, we believe that our conclusions are also accurate for small numbers of quanta. } Closely related metrics have been previously studied in \cite{Hotta:1992qy,Sfetsos:1994xa,Cai:1999dz,Cornalba:2006xk}, following the original Schwarzschild analysis of \cite{Dray:1984ha,Dray:1985yt}. We will construct the geometry by gluing a BTZ solution of mass $M$ to a solution of mass $M+E$ across the null surface $u_w = e^{-R t_w/\ell^2}$. Here, $E$ is the asymptotic energy of the perturbation, which we will take to be very small compared to $M$.

\begin{figure}[ht]
\begin{center}
\includegraphics[scale = 0.7]{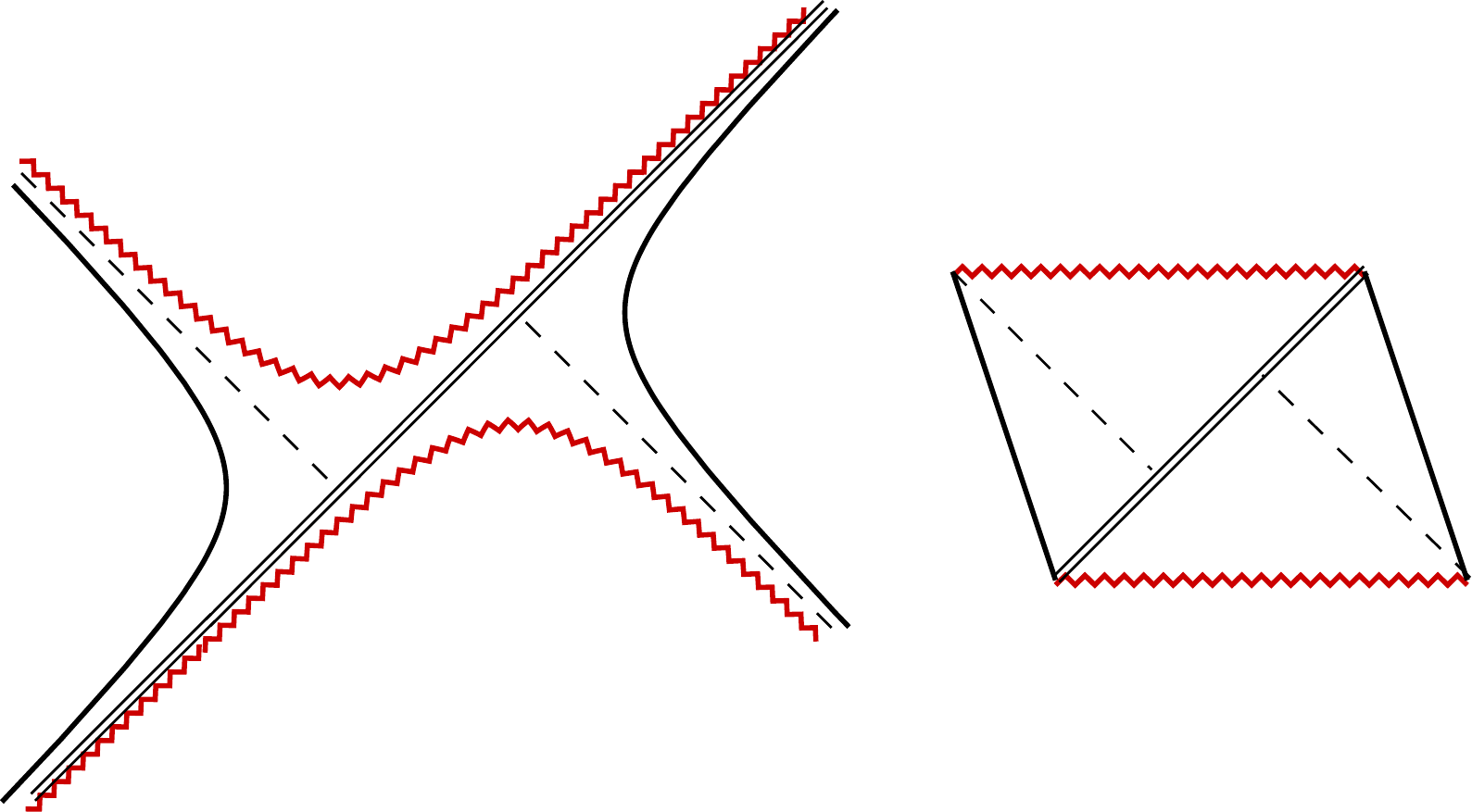}
\caption{The Kruskal and Penrose diagrams for the geometry with a shock wave from the left, represented by the double line. The dashed $v=0$ and $\tilde{v}=0$ horizons miss by an amount $\alpha$.}
\label{oneSideFig}
\end{center}
\end{figure}

We will choose coordinates $u,v$ to the right (past) of the shell and $	\tilde{u},\tilde{v}$ to the left (future), so that the metric is always of the Kruskal form \eqref{kruskal}. Because of the increase in mass, the radius is $R$ to the right and $\tilde{R} = \sqrt{\frac{M+E}{M}}R$ to the left. We will fix a relative boost ambiguity in the relation between $u,v$ and $\tilde{u},\tilde{v}$ by requiring the time coordinate $t$ to flow continuously at the boundary. This determines the location of the shell in terms of the tilded coordinates as $\tilde{u}_w = e^{-\tilde{R}t_w/\ell^2}$. The other matching condition is the requirement that the radius of the $S^{1}$ be continuous across the shell. Inspecting the metric \eqref{kruskal}, we find the condition
\be
\tilde{R}\,\frac{1-\tilde{u}_w\tilde{v}}{1+\tilde{u}_w\tilde{v}} = R \,\frac{1-u_wv}{1+u_wv}\label{match}.
\ee
For small $E/M$, the solution is a simple shift
\be
\tilde{v} = v + \alpha \ , \hspace{30pt} \alpha \equiv \frac{E}{4M}e^{Rt_w/\ell^2} \label{lambda}.
\ee
This matching condition is exact if we take $E/M\to 0$ and $t_w\to \infty$ with $\alpha$ fixed. In this limit, which is relevant for the small but early perturbations we wish to consider, $\tilde{R} = R$, and the metric can be written
\be
ds^2 = \frac{-4 \ell^2 du dv + R^2\left[1 - u(v + \alpha\theta(u))\right]^2d\phi^2}{\left[1+u(v+\alpha \theta(u))\right]^2}.
\ee
The corresponding geometry is shown in Fig.~\ref{oneSideFig}. For computations, it is sometimes useful to use discontinuous coordinates $U = u$, $V = v + \alpha\theta(u)$, so that the metric takes a more standard shock wave form
\be
ds^2 = \frac{-4 \ell^2 dU dV + 4\ell^2\alpha \delta(U)dU^2 + R^2(1-UV)^2d\phi^2}{(1+UV)^2}.
\ee
Either way, the geometry of the patched metric is continuous but its first derivatives are not: there is an impulsive curvature at the location of the shell. One can check that the Einstein equations imply a stress tensor 
\be
T_{uu} = \frac{\alpha}{4\pi G_N} \delta(u),
\ee
corresponding to a shell of null particles symmetrically distributed on the horizon.

\subsection{Geodesics}\label{geodesics}
Since we can boost to a frame in which the shock wave has very little stress energy, the patched solutions described above do not give rise to any large local invariants. The scalar curvature, for example, is regular at $u = 0$. However, there are large {\it nonlocal} invariants that distinguish the shock wave geometry from unperturbed BTZ. Geodesic distance, which we will relate holographically to field theory quantities in \S~\ref{mi} and \S~\ref{corr}, is an important example of such an invariant.

Let us consider a geodesic connecting a point at Killing time $t_L$ on the left boundary with a point at time $t_R$ on the right boundary. We will take both points to be located at the same value of $\phi$. Any real geodesic between them will pass through the shock at $u = 0$ at some value of $v$. We can use the embedding coordinates \eqref{dist} to compute the distance, $d_1$, from the left boundary to this intermediate point and, $d_2$, the distance from the intermediate point to the right boundary:
\begin{align}
\cosh \frac{d_1}{\ell} &= \frac{r}{R} + \frac{1}{R}\sqrt{r^2-R^2} \, e^{-R t_L/\ell^2}(v+\alpha)\\
\cosh \frac{d_2}{\ell} &= \frac{r}{R} - \frac{1}{R}\sqrt{r^2-R^2} \,e^{-Rt_R/\ell^2}v.
\end{align}
To find the total geodesic distance, we extremize $d_1+d_2$ over $v$. For large $r$, the result is
\be
\frac{d}{\ell} = 2\log\frac{2r}{R} + 2\log\Big[\cosh \frac{R}{2\ell^2}(t_R-t_L) + \frac{\alpha}{2}e^{ -R(t_L+t_R)/2\ell^2}\Big].\label{one}
\ee
Setting $\alpha = 0$, we recover the distance in the unperturbed BTZ geometry. The contribution of $\alpha$ represents an increase in this distance due to the shock wave.

It is clear from Eq.~\eqref{one} that the impact of the shock wave on the geodesic distance is insignificant if $t_L+t_R$ is sufficiently large. Indeed, if $t_L\sim t_w$ and $t_R\sim t_w$, then the frame in the bulk defined by the geodesic approximately agrees with the frame natural for the infalling shell. Clearly, the effect on the geometry caused by adding a few quanta should be negligible in this frame, and we don't expect a significant change in the geodesic distance. However, if we fix $t_L = t_R = 0$ and take $Rt_w \gg \ell^2$, then there is a large relative boost between the frame of the quanta and the frame of the geodesic. In the frame of the geodesic, we have highly blueshifted quanta that significantly increase the distance.

We will also record the geodesic distance between two equal-time points on the same boundary, with angular separation $\phi$. This is unaffected by the shock wave, and is given at large $r$ by
\be
\frac{d}{\ell} = 2\log \frac{2r}{R} + 2\log \sinh \frac{R\phi}{2\ell}.\label{three}
\ee

\subsection{Mutual information}\label{mi}
So far in this section, we have constructed the bulk dual to the mildly perturbed thermofield double state. We will now use this geometrical data to understand the behavior of correlations between regions $A\subset L$ and $B\subset R$ in the two CFTs. One useful measure of correlation is the mutual information $I(A;B) = S_A + S_B - S_{A\cup B}$. Employing the RT proposal \cite{Ryu:2006bv} and its time-dependent extension \cite{Hubeny:2007xt}, we can compute the entropy $S_\Omega$ of the density matrix associated to a boundary region $\Omega$ as $A_{min}/4G_N$, where $A_{min}$ is the area of the smallest extremal codimension-two bulk surface that shares a boundary with $\Omega$.\footnote{More precisely, this expression gives the contribution proportional to $N^2$ in the entropy. There may be numerically large but subleading terms, as well as finite $\lambda$ corrections. The RT prescription also requires that the bulk surface must be homologous to $\Omega$.} In a 2+1 dimensional bulk, extremal codimension-two surfaces are geodesics, and the ``area'' is the length of the geodesic.

Following \cite{Morrison:2012iz,Hartman:2013qma}, we will consider a spatial region at $t = 0$ consisting of two disconnected components, $A\subset L$ in the left asymptotic region, and $B\subset R$ in the right asymptotic region. For simplicity, we will take them to be of equal angular size $\phi < \pi$, and we will center them at the same angular location on their respective boundaries. The only subtlety in the calculation arises from the fact that a given spatial region can be bounded by different extremal surfaces. RT instruct us to use the one of minimal area.

First, let us consider $S_A$, or equivalently $S_B$. There are two choices of extremal surface. The first choice is a geodesic that connects the endpoints of the $A$ interval. The other choice is a geodesic that connects one endpoint to the image of the other by the BTZ identification, plus a contribution from the horizon of the black hole required by the RT homology condition. When $\phi < \pi$, the former always has smaller area, and we use \eqref{three} to obtain
\be
S_A = S_B = \frac{\ell}{4G_N}\left(2\log\frac{2r}{R} + 2\log \sinh \frac{R\phi}{2\ell}\right).
\ee

Next, consider $S_{A\cup B}$. When $\phi < \pi$, we have two possible choices of extremal surface. First, we have the union of the two geodesics used to compute $S_A$ and $S_B$. This gives $S_{A\cup B}^{(1)} = S_A + S_B$. Second, we have a pair of geodesics connecting the endpoints of $A$ to the endpoints of $B$. Using \eqref{one}, we find that the second gives
\be
S_{A\cup B}^{(2)} = \frac{\ell}{G_N}\left[\log \frac{2r}{R} + \log\Big(1 + \frac{\alpha}{2}\Big)\right].
\ee
For small regions with $\sinh \frac{R\phi}{2\ell}  <1$, we have $S^{(1)}_{A\cup B} < S^{(2)}_{A\cup B}$, so that $I(A;B)=0$ for all values of $\alpha$ \cite{Morrison:2012iz}. However, for larger regions, $S^{(2)}$ wins for sufficiently small $\alpha$, and we find positive mutual information. Substituting for $\alpha$ using \eqref{lambda}, and rewriting $M$ and $R$ in terms of the Bekenstein-Hawking entropy $S$ and the inverse temperature $\beta$, we obtain
\be
I(A;B) = \frac{\ell}{G_N}\left[\log \sinh \frac{\pi\phi\ell}{\beta} - \log\Big(1 + \frac{E\beta}{4S}e^{2\pi t_w/\beta}\Big)\right].
\ee
This mutual information is a decreasing function of $t_w$. For high temperature, $I$ reaches zero when $t_w$ is equal to
\be
t_{*}(\phi) = \frac{\phi \ell}{2} + \frac{\beta}{2\pi}\log \frac{2 S}{\beta E}.
\ee
When  the string coupling $g_s$, $ \sim 1/N$ in a large $N$ gauge theory is small, so $S \sim N^2$ is large, and $E$ assumes its smallest reasonable value $E \sim T = 1/\beta$ then 
\be
t_* =  \frac{\beta}{2\pi}\log  S.
\ee
as announced in the Introduction.

Similar formulas can be obtained for the case where $\phi > \pi$. There, the mutual information reaches a floor with a finite positive value, rather than zero. One can check that the mutual information between regions with $\phi = \pi$ takes the longest to relax.

\subsection{Correlation functions}\label{corr}
Compared to mutual information, two point functions are a very crude measure of correlation.\footnote{The mutual information is lower-bounded by two-point correlation functions of bounded operators. See e.g. \cite{wolf2008area}.} However, the effect of scrambling on local entanglement is not subtle, and we saw in the spin system that two point functions and mutual information have a qualitatively similar response to a perturbation of the thermofield double state. In this section, we will use the shock wave geometry to obtain an understanding of this response, using the approximation of free field theory on the perturbed background. We first observe that we are interested in computing the following matrix element:
\be
\langle \varphi_L \varphi_R\rangle_W \equiv \frac{ \langle \Psi | W^{\dagger} \varphi_L \varphi_R W | \Psi \rangle}{\langle \Psi | W^{\dagger} W | \Psi \rangle},\label{expval}
\ee
where $W$ is an operator on the left boundary that creates a few particles at a time $t_w$ in the past, and $\varphi_L, \varphi_R$ are the field operators in the L and R theories being correlated, at time $t=0$. $W$ is assumed to have no one-point function in the thermofield double state.

For geometries with a real Euclidean continuation, such as the unperturbed BTZ metric, spacelike correlation functions (in the associated Euclidean vacuum) of CFT operators dual to heavy bulk fields of mass $m$ can reliably be related to the (renormalized) geodesic distance as
\be
\langle \varphi(x)\varphi(y)\rangle \sim e^{-m d(x,y)}.\label{geo}
\ee
This fact has been previously exploited for the purposes of studying black hole interiors \cite{Balasubramanian:1999zv,Louko:2000tp,Kraus:2002iv,Fidkowski:2003nf,Kaplan:2004qe,Festuccia:2005pi}.\footnote{There has been some discussion about whether such two point functions actually diagnose behind-the-horizon physics. The following analysis shows that the two point function is directly sensitive to dynamics that is extremely difficult to interpret solely in terms of supergravity degrees of freedom outside the horizon.} The BTZ shock wave metric is nonanalytic, and analytic approximations do not have real Euclidean continuations. However, in the regime where the shock wave is a small perturbation of the metric, we expect that the the saddle point represented by the perturbed geodesic continues to give the dominant contribution to the two point function, and we can estimate two point functions in the shock wave background using spacelike geodesics that pass through the black hole interior. In fact, an exact calculation of the free field two point function in the BTZ shock wave background has been previously carried out in \cite{Cornalba:2006xk}, and matches our geodesic estimates below up to expected multiplicative corrections of order $1/m\ell$.\footnote{In making the comparison, note that the shift function $h$ in \cite{Cornalba:2006xk} should be identified with twice our shift in $v$.}

Let us therefore proceed to use Eq.~\eqref{geo} to estimate correlation functions. We will focus on the correlator with $t_L = t_R = \phi_R = \phi_L = 0$, and study the dependence on $t_w$. Using the geodesic distance Eq.~\eqref{one}, and subtracting the UV-divergent first term, we obtain the expression
\be
\langle \varphi_L\varphi_R \rangle_W \sim \left(\frac{1}{1 + \frac{E}{8M}e^{Rt_w/\ell^2}}\right)^{2m\ell}.
\ee
This correlator is unaffected by the perturbation until $t_w$ becomes of order $t_*$. For larger $t_w$, the correlator tends to zero exponentially as we make the perturbation earlier, as $e^{-\Delta R(t_w-t_*)/\ell^2}$. 
We emphasize that the value of $t_w$ at which the correlators start to be significantly affected is the same value that we obtained by studying the RT prescription for mutual information. This is not surprising, since the two quantities are determined by the same geodesic data. 
However, there is a significant difference: the mutual information has a sharp feature where it becomes zero shortly after $t_*$, whereas correlators computed in the geodesic approximation merely start to exponentially decay. This discrepancy results from the free field approximation to the scalar correlator. We will see in the next section that inelastic interaction effects turn off the correlation function more sharply after $t_*$.

\section{String and Planck scale effects}\label{speffects}

The analysis of the previous section relies on Einstein gravity.  But, as noted above, a single thermal quantum released at time $t_w$ carries enormous energies in the rest frame of the $t=0$ slice, $E_p \sim \frac{1}{\ell} e^{Rt_w/\ell^2}$.   When $t_w$ is large enough this energy can exceed string or even Planck scales\footnote{We thank Eva Silverstein for valuable discussions about the significance of this situation.}.   The effect on the mutual information comes from string- or Planck-suppressed corrections to the RT formula.  These corrections are not completely understood (but see \cite{deBoer:2011wk,Hung:2011xb}) and so it is difficult to evaluate their effect in the shock wave background.   But we have seen above that the two point correlation function diagnoses similar information.  So we will try to assess in a qualitative way the effect of such corrections on the two point function (\ref{expval}).
\footnote{ We might consider the spacelike geodesic method of calculating the correlation function.   The worldline action should in general contain terms with higher derivatives of the coordinates with respect to proper time, possibly multiplied by curvatures.   These would be multiplied by the appropriate powers of the string mass $m_s$ and the Planck mass $m_p $.  In 3 dimensions these  are related by $m_s \sim g_s^2 m_p$ where $g_s$ is the string coupling.  Here $G_N \sim 1/m_p$.   The results of the previous section show that the high energy in the shock wave causes large derivatives with respect to proper time along the geodesic world line.  These would seem to cause the $1/m_s$ suppressed terms to become order one at times $t_w$ far smaller than  $t_*$.   Because there are no powers of $g_s$ involved,  the time when this would happen would be much sooner,  of order $\log{m_s/T}$ rather than $\log{m_p/T} \sim \log{R/G_N} \sim \log{S}$.  The following remarks about flat-space scattering show that this argument is incorrect.} 
We are interested in the time $t_w$ when this quantity starts to differ substantially from the two point function $\langle \Psi| \varphi_L \varphi_R |\Psi \rangle$.

The expectation value (\ref{expval}) computes spacelike correlations in the state $W|\Psi\rangle$, not scattering information.   So it is difficult to evaluate it in an S-matrix theory like flat space perturbative string theory.  Nonetheless AdS/CFT teaches us that this quantity is well posed in quantum gravity, so there should be some way of understanding it in the region where perturbative string theory is valid. This seems technically difficult, even in BTZ, but some insights might be gained from a string scattering calculation in pure AdS. The methods of \cite{Giveon:1998ns,Brower:2006ea}  could be helpful.

On the other hand, in a situation like the shock where interactions are localized, if we know the spatial correlations at a given time then we can propagate them forward using scattering data.  So we expect that when scattering is weak the change of spatial correlations will be small.  Concretely,  flat space field theory and Einstein gravity calculations in AdS/CFT \cite{Cornalba:2006xk} indicate that when scattering is weak the disturbance of spacelike correlations is also weak.    So we proceed by estimating the strength of flat space string scattering in the relevant energy and coupling regime. 

 The basic features of closed string scattering in flat space in the region of interest are discussed in \cite{Amati:1987wq,Horowitz,Veneziano:2004er,Giddings:2007bw,Hofman:2008ar}.   The largest scattering amplitude occurs in the Regge region, large Mandelstam $s$ and fixed $t$ (as opposed to to the highly suppressed fixed angle region).   Here  the amplitude is small when the dimensionless quantity $\epsilon =g_s^2  s l_s^2$ is small. Putting in $s = E_pT$, appropriate for a thermal quantum sourced by $\varphi_R$, one finds that $\epsilon$ becomes of order one at a time $t_\epsilon$ somewhat prior to $t_*$, by an additive $S$ independent  amount proportional to $\beta\log \ell/l_s$.\footnote{In $D>4$ dimensions.}   So we find that flat-space stringy effects would not change the $\log S$ dependence of the time $t_*$  at which the correlator starts decreasing.

Stringy effects do become important, however. The phase shift obtained from the tree level Virasoro-Shapiro scattering amplitude at large $s$, as a function of impact parameter $b$, agrees with the result of Einstein gravity down to a value $b \sim b_I$ where 
\be
b_I = l_s \sqrt{\log s l_s^2}
\ee
 describes the famous logarithmic spreading of strings at high energy.  For $b < b_I$, there are substantial corrections to the Einstein gravity calculation of the elastic part of the phase shift, summarized by a metric with a transverse profile of size $b_I$ that grows logarithmically with $s$. There are also inelastic processes, that give an imaginary part to the phase shift. The magnitude of the imaginary part of the phase shift is suppressed relative to the real part by $(l_s/b)^2$.
 
 We now turn to scattering\footnote{By ``scattering" behind the horizon we mean an off shell process that resembles scattering with a finite energy and momentum  resolution.}  in the black hole, whose characteristic lengths are the horizon size $R$, the curvature length $\ell$ and the geodesic time from the horizon to the singularity, $\ell$.   The flat space results above are easily applicable only if these lengths are larger than the scales relevant to the flat-space string scattering problem.  This not the case for $t \sim t_*$.
 
 String spreading causes the string to expand in directions transverse to its motion.  Naively it covers the horizon $b_I/R$ times which is roughly  $\sqrt{\log (\epsilon/g_s^2)}$.   Interaction effects should be at most $\epsilon\sqrt{ \log{(\epsilon/g_s^2)}}$, which give a log log correction to $t_*$, which  we ignore.\footnote{On a target space torus, this mild enhancement is completely absent, and therefore may not be present in the black hole problem either.}
 
So far, we have assumed that the scattering takes place far from the singularity and that the string spreading is purely transverse. This may not be the case. If the string spreads significantly in the longitudinal directions,\footnote{Longitudinal spreading in the string ground state has been computed in light-cone gauge in Ref.~\cite{Susskind:1993aa}. This is a large effect, but it appears to be gauge-dependent \cite{Polchinski:1995ta} and we are unsure of its significance to our setup.} the singularity may become important. For this reason, we are unable to reliably exclude the possibility that singularity effects might dramatically enhance the scattering rate. This could have the effect of making $t_*$ much shorter than $\beta\log S$.

For $t_w>t_\epsilon$, several effects come into play. First, there is the increasing effect of the Einstein gravity scattering, which becomes of order one at the time $t_* = \frac{\beta}{2\pi} \log {S}$. In the string scattering problem, this is a purely elastic effect, and should be accurately captured by the homogeneous metric discussed in \S~\ref{holographic}. The inelastic phase shift is suppressed by $l_s^2/\ell^2$, and becomes important at a time $t_w \sim t_* + (const.)\log \ell/l_s$.  This causes the correlator to decay schematically  like $\exp{(-s)}$.    At even larger values of $t_w$, and correspondingly larger energies, a variety of inelastic nonperturbative effects should occur. For instance,  black holes may form.  A rough estimate suggests this occurs when the Schwarzschild radius $R_S$, $R_S^{D-3} \sim G_N \sqrt {s}$ becomes of order $R$.  This occurs at a time $2 t_*$.

Although our estimates have not been conclusive, it is clear that there is an interesting connection between high energy scattering in the black hole background and sensitive dependence on initial conditions in the boundary field theory.   This interplay deserves further attention.

\section{Discussion}\label{discussion}
In the context of Einstein gravity, we have exhibited a bulk holographic  dual to the sensitive dependence on initial conditions in the boundary field theory. Small perturbations at early times create highly blueshifted shock waves that disrupt measures of correlation between the L and R field theories.  The original gravitational interpretation of scrambling as charge spreading on the horizon \cite{Sekino:2008he} is very much in the spirit of our calculation. In particular, the large boost is the source of the logarithmic time dependence. The similarity of the bulk calculations suggests a relation between sensitive dependence on initial conditions and scrambling, and it would be interesting to understand the connection further. 

The shock wave solutions we have used in this paper correspond to boundary sources that are carefully constructed so that all particles launched from the boundary fall into the black hole at the same time.  This allows an exact analytic treatment of the nonlinear general relativity effects at large boost.   A simple local boundary perturbation of the type familiar in field theory would source particles that would fall into the black hole over a band of times, with the probability of staying outside of the black hole decreasing exponentially with the time after the perturbation as $\exp{(-Rt)}$.  In this more general situation each particle still blue shifts after it falls into the black hole and the shock wave metric gives an accurate picture of the disturbance of correlation.  But there are some situations where this spread of infall times becomes important, as we will now discuss.

The observations of the previous section identify inelastic effects that make the correlator behave schematically like $\exp(- s) \sim \exp(-e^t)$.   This is an extraordinarily rapid turnoff, dropping to almost zero at $t \sim t_*$, but is in keeping with the expectations from random dynamics, as in appendix \ref{random}. But because of the spread in infall times we do not actually expect the correlator to go to zero so rapidly.  In the CFT, this corresponds to some amplitude for the perturbation to remain in the ultraviolet degrees of freedom for some time and not to touch off scrambling. If we fold the  $\exp{(-Rt)}$ spread against the $\exp{(-e^t)}$ turnoff  we expect to recover an ordinary exponential decay of the $\langle \varphi_L\varphi_R\rangle_W$ correlator as a function of $t_w$.   The double exponential effect should only leave a subtle imprint, albeit an interesting one.

The shock wave solutions do not display any of the hydrodynamical effects in same side correlators that have been extensively explored in AdS/CFT calculations.  These depend on the nontrivial field profiles connected to the spread in infall times.  As usual the decay of quasinormal modes  and the related hydrodynamical  dissipation are related to the infall of particles through the horizon.  \footnote{The perturbation at $t_w$ can also affect conserved quantities, such as the energy. This will give rise to small but non-decaying terms in the correlation function.}

The interplay of hydrodynamical behavior, in particular diffusive spreading \cite{Susskind:2011ap},  and the scrambling behavior discussed here raises a number of interesting questions for further study.   In particular it would  interesting to study  the spatial propagation of the disturbance of correlations by analyzing the appropriate localized gravity solutions, in contrast to the spherically symmetric perturbations discussed in this paper.    We give a set of such solutions in appendix \ref{localized} but they are adjusted to not give a spread in infall times and so are too specialized to give full insight into this problem.

Although section \ref{holographic} focused on the three-dimensional BTZ geometry, one can consider similar perturbations to higher dimensional black holes. We give a preliminary analysis in appendix \ref{higherd}, where we find that the leading dependence of $t_*$ is universal,  $t_* = \frac{\beta}{2\pi}\log S$. 

Finally we turn to firewalls. The driving force behind the firewall proposal of \cite{Almheiri:2012rt} is a conflict between chaos and specific entanglement \cite{AMPSS}. Although our work is closely related to this issue, and to its recent treatment by Maldacena and Susskind \cite{MaldacenaSusskind}, we are not able to offer any decisive insight. However, we will make a few comments. 

{\bf 1.} Our results provide a new example of an emerging pattern: after a scrambling time, there do not seem to be any simple probes of the behind-the-horizon region. \footnote{But see \cite{Horowitz:2009wm}.} The RT surfaces disconnect, and the correlator goes to zero. 

\begin{figure}[ht]
\begin{center}
\includegraphics[scale = 1]{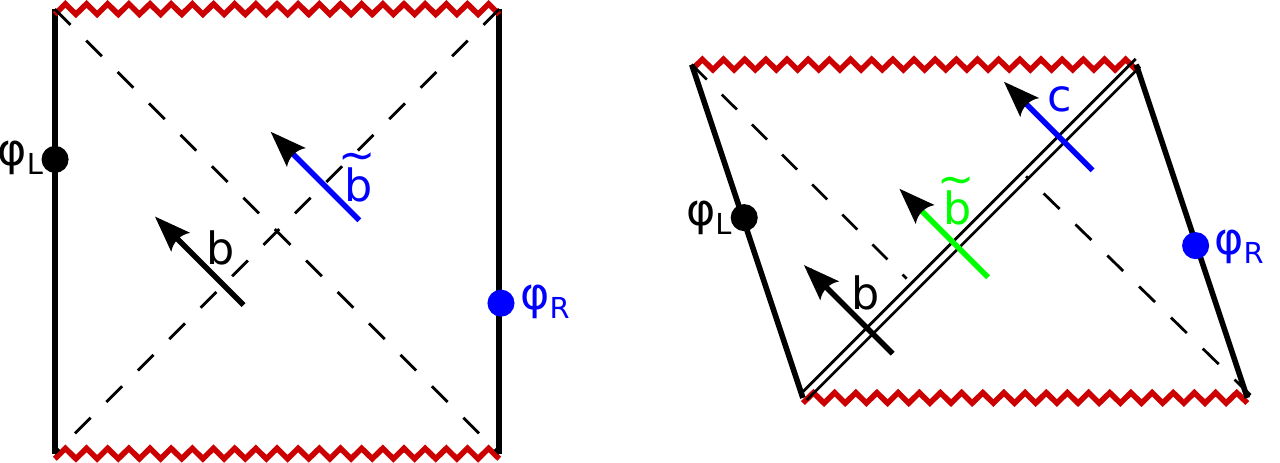}
\caption{In the unperturbed BTZ geometry (left), a smooth horizon requires the black mode on the left to be highly entangled with the blue mode on the right. By contrast, in the shock wave geometry (right) the black and blue modes are far apart and unentangled. Instead, the black mode is entangled with the green mode coming out of the white hole. The arguments of \cite{Leichenauer:2013kaa} suggest that the green mode may be complicated in the CFT.}
\label{modes}
\end{center}
\end{figure}

{\bf 2.} The shock wave geometry defeats a naive argument for firewalls. Smoothness of the left horizon requires entanglement between modes $b$ and $\tilde{b}$ shown in Fig.~\ref{modes}. In the unperturbed geometry, $b$ and $\tilde{b}$ are related to smeared CFT operators $\varphi_L$ and $\varphi_R$ \cite{Bena,HKLL,HMPS}, so the bulk correlation $\langle b \tilde{b}\rangle$ can be viewed as arising from the CFT correlation $\langle\Psi| \varphi_L\varphi_R|\Psi\rangle$ characteristic of the thermofield double state $|\Psi\rangle$. One might worry that the smallness of $\langle\Psi|W^\dagger \varphi_L\varphi_RW|\Psi\rangle$ implies de-correlation of $b$ and $\tilde{b}$ and a firewall in the state $W|\Psi\rangle$.

The geometry shown in Fig.~\ref{modes} gives an alternate explanation.\footnote{A similar situation arises if we consider simultaneous forward time evolution of both CFTs \cite{Hartman:2013qma}, and related comments were made by those authors.} Although $b = \varphi_L$ holds in any geometry that approximates AdS-Schwarzschild outside the horizon, the relationship $\tilde{b} = \varphi_R$ is not valid in the shock wave metric. Indeed, it is clear from Fig.~\ref{modes} that $\varphi_R$ represents a mode $c$ that is far from $b$ and therefore naturally uncorrelated with it. 

{\bf 3.} A stronger argument for firewalls in the state $W|\Psi\rangle$ can be made \cite{AMPSS}. If we assume (i) that the perturbation at sufficiently early $t_w$ acts like a random unitary,\footnote{In this discussion we will abuse notation and refer to a random unitary that approximately commutes with the Hamiltonian as a ``random" unitary.} (ii) that $\tilde{b}$ can always be represented by the same (perhaps very complicated) linear operator, and (iii) that this operator commutes with $b$, then the  counting arguments of AMPSS imply a firewall on both the left and right horizons. To make this argument, one considers the operator $e^{i \theta N_b}$, where $N_b$ is the number operator. This rotates the phase of $\langle b\tilde{b}\rangle$, but leaves the ensemble generated by random unitaries invariant, so we conclude that the ensemble average of $\langle b \tilde{b}\rangle$ must be zero. 

We are unable to determine which, if any, of (i-iii) should be relaxed, but we note that the CFT representation of the green $ \tilde{b}$ mode emerging from the white hole (Fig.~\ref{modes}) is rather mysterious. At higher energies (earlier $t_w$), when the evolution across the shock can no longer be described as a shift, the correct description of the mode $\tilde{b}$ becomes even less clear.

{\bf 4.} The right horizon is not smooth, and the shock would affect an observer falling in from that side \cite{VanRaamsdonk}. In the regime where the shock wave metric is an accurate description, the observer's world line will be abruptly shifted over and the proper time before he hits the singularity reduced. At higher energies, the infaller will experience a painful inelastic collision. Note, however, that for fixed $t_w$, the strength of all such effects decreases as we make the infall time $t_R$ later.  In the regime where Einstein gravity is valid the entanglement of high energy modes is unaffected.  On the other hand, for fixed $t_R$,  we can always make the experience extremely painful by making $t_w$ earlier and earlier.  This suggests a connection between further increasing chaos and the more complete disruption of smooth geometry.   It is clear that this shock wave has many of the attributes of a firewall.\footnote{One might have thought that by making an early perturbation in both CFTs one might have created shock waves on both horizons.  At least for spherical shock waves this is not the case.  The future horizons in the resulting geometry are well beyond the location of the collision and do not coincide with the shock waves \cite{Dray:1985yt}.}

{\bf 5.} Finally, if ``real'' AMPS firewalls form in this system before the scrambling time,  then  our bulk calculations would very likely be inaccurate statements about the CFT dynamics.  We view this as a feature, not a bug.  CFT quantities that are straightforward to formulate (albeit not to calculate!)  would differ from expectations.  

\vspace{1em}
\noindent {\bf Note Added:}

\noindent
After this paper was completed the very interesting paper \cite{Liu:2013iza} appeared which also studies the evolution of entanglement in shock wave geometries. 

\section*{Acknowledgements}
We are grateful to Steve Giddings, Don Marolf,  Joe Polchinski, Eva Silverstein and Lenny Susskind for helpful discussions. We also thank Juan Maldacena and Lenny Susskind for sharing a draft of their paper before publication.  Our work is supported in part by the Stanford Institute for Theoretical Physics and NSF Grant 0756174. We both acknowledge the hospitality of the Kavli Institute for Theoretical Physics and NSF Grant PHY11-25915 for support.   DS was also supported by a KITP Graduate Fellowship, and by the NSF GRF program.  SS is grateful for support at KITP made possible by the Simons Foundation.

\appendix

\section{Haar scrambling}\label{random}
In the main text of the paper, we've considered the effect of an operator $\mathcal{O}_L(t_w) = e^{-it_wH} \mathcal{O}_Le^{it_wH}$ on the the entanglements between local subsystems $A\subset L$ and $B\subset R$ in a thermofield double state $|\Psi\rangle_{LR}$. If the Hamiltonian is sufficiently chaotic, and we take very large values of $t_w$, we might model such a perturbation as a random unitary matrix, so that the perturbed thermofield double state is
\be
|\Psi'\rangle = \frac{1}{|L|^{1/2}}\sum_{n,m = 1}^{|L|}U_{nm}|m\rangle_L|n\rangle_R.\label{state}
\ee
In this appendix, we will study the mutual information $I(A;B)$ and correlation functions in this state, using the tool of Haar integrals. The result will not be surprising to the reader familiar with Page's analysis of random states \cite{Page:1993df}. However, our setup is not identical to that of Page, and we will include the discussion for completeness, following the computationally efficient norm approach of \cite{Hayden:2007cs}.\footnote{This approach has the benefit of emphasizing that Haar randomness is not essential to the calculation, and that a 2-design would lead to identical results.}

Specifically, in order to study the mutual information $I(A;B)$ in this state, we will consider the distance $d_1 = \|\rho_{AB} - \rho_A\otimes \rho_B\|_1$, where the 1-norm of a matrix $M$ is defined as $\|M\|_1 = tr[\sqrt{M^\dagger M}]$. For $d_1 \le 1/e$, this quantity lower-bounds the mutual information via the Pinsker inequality, and upper-bounds it via the Fannes inequality \cite{fannes}:
\be
\frac{1}{2}d_1^2\le I(A,B) \le d_1\log|A| - d_1\log d_1.\label{Fannes}
\ee
Unfortunately, because of the square-root, the distance $d_1$ is difficult to average over $U$, so we will use a further inequality (derived from Cauchy-Schwarz applied to the eigenvalues), that $\|M\|_1 \le \sqrt{\text{rank}\,M}\|M\|_2$, where the 2-norm is defined as $\|M\|_2 = \sqrt{tr[M^\dagger M]}$. The benefit here is that we can compute the average over unitaries of the square of the 2-norm exactly. Using the fact that for any $U$, the density matrix for $A$ obtained from $|\Psi\rangle$ is maximally mixed, $\rho_A(U) =\rho_B(U) =  1/|A|$, we compute
\begin{align}
\|\rho_{AB}(U) - \rho_A(U)\otimes \rho_B(U)\|_2^2 &= tr[(\rho_{AB}(U) - 1/|A|^2)^2] \\
&= tr[\rho_{AB}(U)^2] - \frac{2}{|A|^2}tr[\rho_{AB}] + \frac{1}{|A|^4}tr[1] \\
&= tr[\rho_{AB}(U)^2] - \frac{1}{|A|^2}.\label{comp}
\end{align}

We would now like to take the expectation value over $U$ of this quantity, using the Haar measure. The most direct way to do this computation is to break up the $m$ and $n$ indices of $U_{mn}$ into $m \rightarrow (i,I)$, where $i$ runs over the $A$ Hilbert space, and $I$ runs over the Hilbert space of $A^c$, the tensor complement of $A$ in $L$. The operator $U$ is then represented as $U_{iI\,jJ}$, and one can check that 
\be
tr[\rho_{AB}(U)^2] = \frac{1}{|L|^2}\sum_{iIjJi'I'j'J'}U_{iI\,jJ}U^*_{i'I\,j'J}U_{i'I'\,j'J'}U^*_{iI'\,jJ'}.\label{hi}
\ee
We can now take the expectation value using
\begin{align}
\int dU \ U_{i_1 j_1}U_{i_2 j_2}U^*_{i_1' j_1'} U^*_{i_2' j_2'} &= \frac{1}{|L|^2-1}\Big(\delta_{i_1 i_1'}\delta_{i_2 i_2'}\delta_{j_1j_1'}\delta_{j_2j_2'} + \delta_{i_1 i_2'}\delta_{i_2 i_1'}\delta_{j_1j_2'}\delta_{j_2j_1'} \Big) \\ &\hspace{40pt} -\frac{1}{|L|(|L|^2-1)}\Big(\delta_{i_1i_1'}\delta_{i_2 i_2'}\delta_{j_1j_2'}\delta_{j_2j_1'} + \delta_{i_1 i_2'}\delta_{i_2 i_1'}\delta_{j_1j_1'}\delta_{j_2j_2'}  \Big).\notag
\end{align}
The terms on the bottom line are subleading and we will drop them, along with the ``1'' in the first line. Summing as in Eq. \eqref{hi}, we find that $tr[\rho_{AB}(U)^2] = |A|^{-2} + |A^c|^{-2}$. Using the convexity of the square root, the Cauchy-Schwarz inequality and Eq. \eqref{comp}, this implies 
\be
\int dU \|\rho_{AB}(U) - \rho_A(U)\otimes \rho_B(U)\|_1 \le \frac{|A|}{|A^c|}.
\ee
If $A$ is less than half of the $L$ system, the one-norm distance is suppressed by a ratio of Hilbert space dimensions. This quantity is exponentially small in, e.g. the number of extra qubits in $A^c$ compared to $A$. Using Eq. \eqref{Fannes}, we can bound the mutual information
\be
\int dU\, I(A,B) \le \frac{|A|}{|A^c|}\log |A^c|.
\ee
The large logarithmic factor is probably an artifact of our shortcut through the 1-norm, but in any case, if $A$ is significantly smaller than half of the total system, the above is exponentially small.

If the $L$ and $R$ systems are composed of qubits, we can also study correlation functions of a spin in the $L$ system and a corresponding spin in the $R$ system. One can bound these correlations using the computation of $d_1$ above, but a direct calculation in the state \eqref{state} is simple enough. Averaging over $U$, one finds that the expected value of the spin-spin correlator is zero, and the rms value is $|L|^{-1}$.

\section{Geometrical generalizations}
\subsection{Higher dimensions}\label{higherd}
Our analysis in this paper was largely restricted to three spacetime dimensions. In this appendix, we will explore the effect of the shock wave for $D$-dimensional AdS black holes. $D$ is the bulk spacetime dimension, i.e. $D = 3$ for BTZ. We will not attempt to compute geodesic distances and RT surfaces. As a simpler proxy, we will estimate how large $t_w$ has to be to make the shift in the $v$ coordinate of order one.\footnote{The main point, that the coefficient of the logarithm in $t_*$ is dimension-independent, should already be clear from the discussion in \S~\ref{speffects}.}

We begin with the metric
\be
ds^2=-f(r)dt^2 + f^{-1}(r)dr^2 + r^2 d\Omega_{D-2}^2\label{r,t}.
\ee
Assuming the existence of a horizon at $r = R$, we pass to Kruskal coordinates:
\begin{align}
ds^2&=-\frac{4 f(r)}{f'(R)^2}e^{-f'(R)r_*(r)}dudv + r^2 d\Omega_{D-2}^2\\
uv &= -e^{f'(R) r_*(r)} \hspace{20pt} u/v = -e^{-f'(R)t},
\end{align}
with $dr_* = f^{-1}dr$ the usual tortoise coordinate. As in \S~\ref{shock}, we add a spherically symmetric null perturbation of asymptotic energy $E \ll M$, at a time $t_w$ in the left asymptotic region. We define coordinates $\tilde{u},\tilde{v}$ to the left of the perturbation, and continue to use $u,v$ to the right. The shell propagates on the surface
\be
\tilde{u}_w = e^{\frac{\tilde{f}'(\tilde{R})}{2}(\tilde{r}_*(\infty) - t_w)} \hspace{20pt} u_w = e^{\frac{f'(R)}{2}(r_*(\infty) - t_w)},\label{u}
\ee
and the matching condition relates $\tilde{v}$ to $v$ via
\be
\tilde{u}_w\tilde{v} = -e^{\tilde{f}'(\tilde{R})\tilde{r}_*(r)} \hspace{20pt} u_wv= -e^{f'(R)r_*(r)}.
\ee
We would like to use these equations to find the shift $\tilde{v} = v + \alpha$, to linear order in $E$ and at large $t_w$. Small $E$ allows us to approximate $\tilde{u}_w = u_w$. Large $t_w$ pushes us to a limit where $r$ approaches $R$, so we can expand $f(r) = f'(R)(r-R)+...$. Evaluating $r_*$, we find $e^{f'(R)r_*(r)} = (r-R)C(r,R)$, where $C$ is smooth and nonzero at $r = R$. To linear order in $E$ and at large $t_w$, we therefore have
\be
\alpha = \frac{E}{u_w} \frac{d}{dM}\Big[(R-r)C(r,R)\Big]\Big|_{r=R} = \frac{E}{u_w}\frac{dR}{dM}C(R,R).
\ee
We can relate $R$ to $S_{BH}$ using the area formula, and use the first law of thermodynamics to evaluate $dR/dM$. Also using $f'(R) = 4\pi/\beta$, we find that $\alpha$ becomes equal to one at time\footnote{Note that the combination $r_*(\infty) - \frac{\beta}{2\pi}\log C(R,R)$ is invariant under additive shifts in the definition of $r_*(r)$.}
\be
t_* = r_*(\infty) + \frac{\beta}{2\pi}\log\left[\frac{(D-2)\Omega_{D-2}}{4C(R,R)}\frac{T}{E}\frac{R^{D-3}}{G_N}\right].
\ee
Fixing $E,R,T$ and taking $G_N \propto N^{-2}$ to zero, we have $t_* \sim \frac{\beta}{2\pi}\log N^2$ in any spacetime dimension.

\subsection{Solutions with localized sources}\label{localized}
Additional insight into the process of de-correlation might be gained by considering solutions with stress energy localized in the angular directions. The interpretation of such solutions is subtle, but we will record their form here. We focus on the case where the shock is produced by a very low-energy perturbation long in the past, so it lies entirely on the right horizon. We assume the infalling source is at the north pole of the $(D-2)$-sphere, and we make the ansatz $\tilde{v} = v + h(\Omega)$. Evaluating the Ricci tensor of the patched metric (see, e.g. appendix A of \cite{Sfetsos:1994xa}) and plugging into the Einstein equations with $T_{uu}\propto \delta(u)\delta^{D-2}(\Omega)$, we find that $h$ must satisfy an equation 
\be
\left[\nabla^2_{S^{D-2}} - \frac{D-2}{2}Rf'(R)\right]h(\Omega) \propto \delta^{D-2}(\Omega).
\ee
For a large $AdS$ black hole with $R\gg \ell$, we have $f'(R) \approx (D-1)R/\ell^2$. The shift $h$ is the Green's function for a very massive field on the sphere, and it decays with angular distance from the north pole as $h\propto e^{-\sqrt{\frac{(D-1)(D-2)}{2}}R\theta/\ell}$. Comparing this to the rate at which the perturbation grows as we push $t_w$ earlier, $e^{f'(R)t_w/2}$, we find that as we increase $t_w$, the level sets of $h$ expand outward with a ``speed of propagation''
\be
v_D = \sqrt{\frac{D-1}{2(D-2)}},
\ee
where $D$ is the spacetime dimension of the AdS space.

\end{document}